\documentclass[12pt,a4]{article}

\usepackage{amsmath}
\usepackage{graphicx}
\usepackage[table]{xcolor}
\usepackage{booktabs}
\usepackage{amssymb,bm}
\usepackage{textcomp}
\usepackage{multirow}
\usepackage{subcaption}
\usepackage{url}

\begin{document}
	
	\title{Short-duration Speaker Verification (SdSV) Challenge 2021: the Challenge Evaluation Plan}
	
	\author{Hossein Zeinali, Kong Aik Lee, Jahangir Alam, Luka\v{s} Burget}
	
	\date{Version 1.0, January 23, 2021}
	
	\maketitle
	
	\section{Introduction}
	
	
	
	
	This document describes the Short-duration Speaker Verification (SdSV) Challenge 2021 which focuses on the analysis and exploration of new ideas for short duration speaker verification. The main purpose of this challenge is to encourage participants on building single but competitive systems, to perform analysis as well as to explore new ideas, such as multi-task learning, unsupervised/self-supervised learning, single-shot learning, disentangled representation learning and so on, for short-duration speaker verification.
	
	Following the success of the SdSV 2020, the SdSV Challenge 2021 features an unusual and difficult task of cross-lingual speaker verification (English vs. Persian), in addition to the text-dependent (TD) and text-independent (TI) tasks. The evaluation dataset used for the challenge is drawn from the recently released multi-purpose DeepMine dataset~\cite{deepmine2018odyssey,deepmine2019asru}. The dataset has three parts and among them, Part 1 is used for TD-SV while Part 3 is for TI-SV.
	
	For access to the challenge description of SdSV Challenge 2020, you can refer to this link: \url{https://sdsvc.github.io/2020/assets/SdSV_Challenge_Evaluation_Plan.pdf}, and more information about it can be found in \url{https://sdsvc.github.io/2020/}.
	
	\section{Tasks Description}
	
	\subsection{Task 1 - Text-Dependent Speaker Verification}
	
	Task 1 of the SdSV Challenge 2021 is defined as speaker verification in text-dependent mode: given a test segment of speech and the target speaker's enrollment data, automatically determine whether a specific phrase and the test segment was spoken by the target speaker. In contrast to text-independent speaker verification, the lexical content of the utterance is also taken into consideration. As such, Task 1 is a twofold verification task in which both the speaker and phrase are verified.
	
	In Task 1, each trial consists of a test segment along with a model identifier which indicates three enrollment utterances and a phrase ID that uttered in the utterances. The system is required to process each trial independently and produce a log-likelihood ratio (LLR) score which combines both speaker and phrase verification scores.
	
	The enrollment and test phrases are drawn from a fixed set of ten phrases consisting of five Persian and five English phrases, respectively. The in-domain training data contains utterances from \textbf{963 speakers}, some of which have only Persian phrases. Model enrollment is done in a phrase and language-dependent way using three utterances for each model. Given the same set of target speakers, Task 1 provides a good basis for analyzing the language factor in text-dependent speaker recognition.
	
	\subsubsection{Trial types}
	
	Given the ground truth, there are four types of trial in a TD-SV task ~\cite{larcher2014text}. The first is Target-Correct (TC) where the target speaker utters the correct pass-phrase. The second is Target-Wrong (TW) where the target speaker utters a wrong pass-phrase. In the same manner, the Imposter-Correct (IC) and Imposter-Wrong (IW) refer to the case where the imposter utters the correct or a wrong pass-phrase. The system should accept the TC trials as the target trial and reject the other three types as non-target (imposture) trials. Notice that the main difference between text-dependent and text-independent speaker verification is considering TW trials as imposter trial while both TC and TW are considered as target trials in the text-independent mode. There are no cross-language and cross-gender trials in Task 1 of the challenge.
	
	\subsubsection{Training condition}
	
	The training condition is defined as the amount of data/resources used to build a Speaker Recognition (SR) system. Same as SdSVC 2020, we adopted a fixed training condition where the system should only be trained using a designated set. The fixed training set consists of the following:
	
	\begin{itemize}
		\item VoxCeleb1
		\item VoxCeleb2
		\item LibriSpeech
		\item Mozilla Common Voice Farsi
		\item DeepMine (Task 1 Train Partition)
	\end{itemize}
	
	The use of other public or private speech data for training is forbidden, while the use of non-speech data for data augmentation purposes is allowed. The in-domain DeepMine training data can be used for any purpose, such as neural network training, LDA or PLDA model training, and score normalization. 
	Unlike SdSVC 2020, this year we provide a \textbf{separate development set} for the challenge. Teams are allowed to use the development set for parameter tuning and evaluation of the systems before submitting them to the leaderboard to save the number of submissions. Any other training such as fusion training on the development set is not allowed because its speakers have overlap with the main evaluation data. Note also, any usage of Task 2 in-domain data and its development set data for this task is not allowed.
	
	
	\subsubsection{Enrollment Condition}
	
	The enrollment is accomplished using three utterances of a specific phrase for each model. We decided to use three utterances for the model enrollment since it is commonly adopted in practice. Note that using enrollment utterances of other models is forbidden, for example, for calculating score normalization parameters (i.e., trials are to be processed independently).
	
	\subsubsection{Test Condition}
	
	Each trial in the evaluation contains a test utterance and a target model. As described above, there are four types of trials in the evaluation and only TC is considered as target and the rest will be considered as imposture. 
	Unlike SdSVC 2020, this year we only have a smaller evaluation set (so the progress set is eliminated) in the Codalab that is used to monitor progress on the leaderboard as well as the final ranks of the participants. Based on this, on the last few days of the challenge deadline, the leaderboard will be hidden.
	
	\subsection{Task 2 - Text-Independent Speaker Verification}
	
	Task 2 of the SdSV Challenge is speaker verification in text-independent mode: given a test segment of speech and the target speaker enrollment data, automatically determine whether the test segment was spoken by the target speaker.
	
	Each trial in this task contains a test segment of speech along with a model identifier which indicates one to several enrollment utterances. The net enrollment speech for each model is distributed between 4 to 180 seconds (after applying an energy-based VAD). The system is required to process each trial independently and produce a log-likelihood ratio (LLR) for each of them.
	
	The in-domain training data in this task contains text-independent Persian utterances from \textbf{588 speakers}. This data can be used for any purpose such as LDA/PLDA, score normalization, training data for neural network, reducing the effect of language for cross-lingual trials, etc.
	
	\subsubsection{Trials}
	
	There are two partitions in this task. The first partition consists of typical text-independent trials where the enrollment and test utterances are from the same language (Persian). The second partition consists of text-independent cross-language trials where the enrollment utterances are in Persian and test utterances are in English. For this partition, the system should reduce the language effects in order to verify the test utterances in a different language. Similar to Task 1, there are no cross-gender trials in Task 2. Note that any further information about test language will not be provided but participants are allowed to train any language identification system to do it if they needed.
	
	\subsubsection{Training condition}
	
	Similar to Task 1, we adopted a fixed training condition where the system should only be trained using a designated set. The available training data is as follow:

	\begin{itemize}
		\item VoxCeleb1
		\item VoxCeleb2
		\item LibriSpeech
		\item Mozilla Common Voice Farsi
		\item DeepMine (Task 2 Train Partition)
	\end{itemize}
	
	The use of other public or private speech data for training is forbidden, while the use of non-speech data for data augmentation purposes is allowed. The in-domain DeepMine training data can be used for any purpose, such as neural network training, LDA or PLDA model training, and score normalization.
    Unlike SdSVC 2020, this year we provide a \textbf{separate development set} for the challenge. Teams are allowed to use the development set for parameter tuning and evaluation of the systems before submitting them to the leaderboard to save the number of submissions. Any other training such as fusion training on the development set is not allowed because its speakers have overlap with the main evaluation data. Note also, any usage of Task 1 in-domain data and its development set data for this task is not allowed.
	
	\subsubsection{Enrollment Condition}
	
	The enrollment data in Task 2 consists of one to several variable-length utterances. The net speech duration for each model is roughly 4 to 180 seconds. Since each enrollment utterance is a complete recording without trimming to a specific duration, the overall duration might not be exactly uniform. Note that using the enrollment utterances from the other models is forbidden, for example, for calculating score normalization parameters.
	
	\subsubsection{Test Condition}
	
	Each trial in the evaluation contains a test utterance and a target model. The duration of the test utterances varies between 1 to 8 seconds. 
	Unlike SdSVC 2020, this year we only have a smaller evaluation set (so the progress set is eliminated) in the Codalab that is used to monitor progress on the leaderboard as well as the final ranks of the participants. Based on this, on the last few days of the challenge deadline, the leaderboard will be hidden.
	
	\subsection{Performance Measurement for Both Tasks}
	
	The main metric for the challenge is normalized minimum Detection Cost Function (DCF) as defined is SRE08. This detection cost function is defined as a weighted sum of miss and false alarm error probabilities:
	\begin{equation*}
	\label{eq.s_norm}
	C_{Det} = C_{Miss} \times P_{Miss\:|\:Target} \times P_{Target} + C_{FalseAlarm} \times P_{FalseAlarm\:|\:NonTarget} \times (1 - P_{Target})\:,
	\end{equation*}
	where $C_{Miss} = 10$, $C_{FalseAlarm} = 1$ and $P_{Target}=0.01$. Based on the parameters, the normalized DCF ($DCF_{norm}$) will be DCF divide by 0.1 as the best cost that could be obtained without processing the input data. In addition to $minDCF_{norm}^{0.01}$, the Equal Error Rate (EER) will be reported.
	
	\subsection{Data Description}
	
	The main data for the challenge is the DeepMine dataset which was collected using crowdsourcing~\cite{lee2015reddots}. Participants in the data collection project installed an Android application and record phrases in the application. The full description of the project and the dataset can be found in~\cite{deepmine2019asru, deepmine2018odyssey}. Bibtex sources for citing the database are provided below easy reference:
	
	\begin{verbatim}
	@inproceedings{deepmine2018odyssey,
	    title={{DeepMine} Speech Processing Database: Text-Dependent and 
	    Independent Speaker Verification and Speech Recognition in 
	    {Persian and English}.},
	    author={Zeinali, Hossein and Sameti, Hossein and Stafylakis, Themos},
	    year=2018,
	    booktitle={Proc. Odyssey 2018 The Speaker and Language Recognition 
	    Workshop},
	    pages={386--392},
	}
	
	@inproceedings{deepmine2019asru,
	    title={A Multi Purpose and Large Scale Speech Corpus in {Persian and 
	    English} for Speaker and Speech Recognition: the {DeepMine} Database},
	    author={Zeinali, Hossein and Burget, Lukas and Cernocky, Jan},
	    year=2019,
	    booktitle={Proc. ASRU 2019 The 2019 IEEE Automatic Speech Recognition 
	    and Understanding Workshop},
	}
	\end{verbatim}

	The database was recorded in realistic environments, which took place in Iran, and the collection protocol was designed to incorporate various kinds of noises in the collection. The main language is Farsi (Persian) while most of the participants also participated in the English partition. Part 1 of the dataset contains five Persian phrases as well as five English phrases that are used in Task 1 of the challenge. The English phrases and transliteration of the Persian phrases are shown in Table~\ref{tbl.phrases}. The phoneme transcription of the phrases will be provided with the data and participants can use them in any way they want. Part 3 of the dataset contains text-independent phrases in Persian and is using in Task 2 of the challenge as the enrollment data.
	
	\begin{table}[tb]
		\renewcommand{\arraystretch}{1.2}
		\caption{\label{tbl.phrases} Phrases in Task1 of the challenge.}
		\vspace{-2mm}
		\centerline
		{
			\setlength\tabcolsep{12pt}
			\begin{tabular}{l | l }
				\toprule
				\midrule
				Id & Phrase \\
				\midrule
				01 & sedaye man neshandahandeye hoviyyate man ast. \\
				02 & sedaye har kas monhaser be fard ast. \\
				03 & hoviyyate man ra ba sedaye man tayid kon. \\
				04 & sedaye man ramze obure man ast. \\
				05 & baniadam azaye yekdigarand. \\
				06 & My voice is my password. \\
				07 & OK Google. \\ 
				08 & Artificial intelligence is for real. \\
				09 & Actions speak louder than words. \\
				10 & There is no such thing as a free lunch. \\
				\midrule
				\bottomrule
			\end{tabular}
		}
	\end{table}
	
	\subsection{Data Organization}
	
	Data will be provided in three separate zip (tar) files for each task. The first zip file contains the in-domain DeepMine training data. The second zip file contains enrollment data, model definition file, and trial file. The last zip file contains test utterances. If all three files will be extracted in a directory, the directory structure is as follow:
	
	\begin{verbatim}
	<base directory>/
	  README.txt
	  docs/
	    model_enrollment.txt
	    train_labels.txt
	    trials.txt
	  wav/
	    enrollment/
	      enr_000000.wav
	      enr_000001.wav
	      ...
	    evaluation/
	      evl_000000.wav
	      evl_000001.wav
	      ...
	    train/
	      trn_000000.wav
	      trn_000001.wav
	      ...
	\end{verbatim}
	
	Note that the zip files for Task 1 and Task 2 should be extracted into separate directories since their contents are partially overlapping. 
	Also, note that for SdSV Challenge 2021, the \textit{trials.txt} file is different from the last year, and lots of trials are eliminated from it (progress set and some trials from the evaluation set of the last year). The new trials file for each task will be provided in the registration confirmation email.
	
	\subsection{Format of Model Enrollment File}
	
	\subsubsection{Task 1 Enrollment File}
	
	The enrollment file of Task 1 is a space-separated five-column text file \texttt{model\_enrollment.txt} located in the \texttt{docs} directory. There is a header line at the beginning of the file. The first record in each line indicates a model-ID, the second record shows the phrase ID which indicates the phrase uttered in the corresponding utterances. The remaining three columns show the enrollment file IDs. There is only one space between two records in each line. The format of the enrollment file is as follow:
	
	\begin{verbatim}
	model-id<SPACE>phrase-id<SPACE>enroll-file-id1<SPACE>enroll-file-id2
	<SPACE>enroll-file-id3<NEWLINE>
	\end{verbatim}
	where \texttt{model-id} is the model identifier, \texttt{phrase-id} is the phrase identifier and \texttt{enroll-file-ids} are the enrollment utterance identifiers.
	
	For example:
	
	\begin{verbatim}
	model-id phrase-id enroll-file-id1 enroll-file-id2 enroll-file-id3
	model_00000 07 enr_007492 enr_023277 enr_012882
	model_00001 02 enr_035341 enr_027674 enr_032835
	model_00002 09 enr_020095 enr_015193 enr_024742
	model_00003 06 enr_032246 enr_014610 enr_014698
	model_00004 09 enr_033841 enr_037127 enr_033859
	\end{verbatim}
	
	\subsubsection{Task 2 Enrollment File}
	
	The enrollment file of Task 2 is a space-separated text file \texttt{model\_enrollment.txt} located in the \texttt{docs} directory. There is a header line at the beginning of the file. There are at least two records in each line. The first record indicates a model ID and the second record shows the first enrollment file. If the model has more than one enrollment files, the rest will follow in the same line. The number of columns in each line may differ depending on the number of enrollment files. There is only one space between two records in each line. The format of the enrollment file is as follow:
	
	\begin{verbatim}
	model-id<SPACE>enroll-file-id1[<SPACE>enroll-file-id2
	[<SPACE>enroll-file-id3...]]<NEWLINE>
	\end{verbatim}
	where \texttt{model-id} is the model identifier and \texttt{enroll-file-ids} are the enrollment utterance identifiers.
	
	For example:
	
	\begin{verbatim}
	model-id enroll-file-ids ...
	model_15002 enr_110254 enr_264593
	model_15005 enr_188426 enr_303480 enr_200614 enr_117624
	model_15006 enr_072239 
	model_15007 enr_248083 enr_316783 enr_088834 
	model_15008 enr_177720 enr_334136 enr_226306 enr_057733 enr_190105
	model_15009 enr_059968 enr_234582 
	model_15011 enr_310490 enr_264156 enr_055518 enr_091529
	\end{verbatim}
	
	\subsection{Format of Trial File}
	
	The trial file for both tasks is a space-separated two-column text file \texttt{trials.txt} located in the \texttt{docs} directory. There is a header line at the beginning of the file. The first record in each line indicates a model ID and the second record indicates an evaluation file ID. There is only one space between two records in each line. The format of the trial file is as follow:
	
	\begin{verbatim}
	model-id<SPACE>evaluation-file-id<NEWLINE>
	\end{verbatim}
	where model-id is the model identifier and evaluation-file-id is the test utterance identifier.
	
	For example:
	
	\begin{verbatim}
	model-id evaluation-file-id
	model_00000 evl_000018
	model_00000 evl_000021
	model_00000 evl_000035
	model_00000 evl_000109
	model_00000 evl_000117
	model_00000 evl_000165
	\end{verbatim}
	
	\section{In-domain Training Set}
	
	\subsection{Task 1 In-domain Data}
	
	As described earlier, the in-domain data for Task 1 consists of utterances from 963 speakers. All training utterances are stored in the \texttt{wav/train} directory. The \texttt{train\_labels.txt} file in the \texttt{docs} directory is a space-separated text file that contains the information for each utterance. Each line in this file contains three columns, where the first column shows \texttt{train-file-id}, the second one indicates the \texttt{speaker-id} and the last one shows \texttt{phrase-id}. There is a header line at the beginning of the file. The format of train label file is as follow:
	
	\begin{verbatim}
	train-file-id<SPACE>speaker-id<SPACE>phrase-id<NEWLINE>
	\end{verbatim}
	where \texttt{rain-file-id} is the train utterance identifier, \texttt{speaker-id} is the speaker label and finally, the \texttt{phrase-id} is the identifier of phrase of each utterance.
	
	For example:
	
	\begin{verbatim}
	trn_000001	spk_000001	09
	trn_000002	spk_000001	09
	trn_000003	spk_000001	09
	trn_000004	spk_000001	09
	trn_000005	spk_000001	09
	trn_000006	spk_000001	09
	trn_000007	spk_000001	09
	\end{verbatim}
	
	\subsection{Task 2 In-domain Data}
	
	The in-domain data for Task 2 consists of utterances from 588 speakers. All training utterances are stored in the \texttt{wav/train} directory. The \texttt{train\_labels.txt} file in docs directory is a space-separated text file that contains the information for each utterance. Each line in this file contains two columns, where the first column shows \texttt{train-file-id} and the second one indicates the \texttt{speaker-id}. There is a header line at the beginning of the file. The format of train label file is as follow:
	
	\begin{verbatim}
	train-file-id<SPACE>speaker-id<NEWLINE>
	\end{verbatim}
	where \texttt{train-file-id} is the train utterance identifier and \texttt{speaker-id} is the speaker label.
	
	For example:
	
	\begin{verbatim}
	trn_101064 spk_001000
	trn_101065 spk_001000
	trn_101066 spk_001000
	trn_101067 spk_001000
	trn_101068 spk_001000
	trn_101069 spk_001000
	trn_101070 spk_001000
	\end{verbatim}
	
	\section{Evaluation Rules and Requirements}
	
	The overall rules are pretty the same as NIST SREs. First of all, participants must follow the data restriction where there is only a fixed training condition. Participants are also required to process the test data according to the following rules and upload results to the challenge website for evaluation. These rules are:
	
	\begin{itemize}
		\item Participants agree to make at least one valid submission for one of the tasks.
		\item Participants agree to process each trial independently. That is, each decision for a trial is to be based only upon the specified test segment and target speaker enrollment data. The use of information about other test segments and/or other target speaker data is not allowed.
		\item Participants agree not to probe the enrollment or test segments via manual/human means such as listening to the data or producing the manual transcript of the speech.
		\item Participants are allowed to use any automatically derived information for training, development, enrollment, or test segments.
		\item Participants are not allowed to use the development set for any training purposes.
		\item Participants may make multiple submissions for each task. Based on the leaderboard results, participants should select the best performing systems, write a proper system description based on the results, and sent it to challenge organizers via email \url{sdsv.challenge@gmail.com}.
	\end{itemize}
	
	\section{Baselines}
	
	There is a common x-vector~\cite{snyder2018x} baseline for both text-dependent and independent tasks. Also, we will prepare the second baseline for Task 1 (text-dependent) during the challenge period. We provide below a brief description of the baselines. The source code for the x-vector baseline can be found on the challenge website.
	
	\subsection{X-Vector Baseline}
	
	The first baseline is a well-known x-vector system for text-independent speaker verification~\cite{snyder2018x}. Here E-TDNN topology~\cite{snyder2019speaker} is used and trained using VoxCeleb1 and VoxCeleb2. After training the network, an LDA with 150 dimensions is applied to the extracted x-vectors and after that, a PLDA with both VoxCeleb datasets is trained. Finally, trials are scored using the PLDA without any score normalization.
	
	\subsection{I-Vector/HMM Baseline}
	
	Because the x-vector baseline does not consider any phrase information and is not optimized for the text-dependent task, the i-vector/HMM method is selected as the second baseline for Task1. The method was proposed in~\cite{zeinali2016deep, zeinali2017hmm} and have achieved very good results on both RSR2015~\cite{larcher2014text} and RedDots~\cite{lee2015reddots} datasets.
	
	In this method, i-vector is used as a fixed dimension representation for each utterance. In contrast to the conventional i-vector/GMM method which uses GMM for aligning the frames to Gaussian components, here monophone HMMs are used as frame alignment strategy. So, the first step is training monophone HMMs using the training data. These models then will be used to extract sufficient statistics and by using them an i-vector extractor is trained. Scoring is done using LDA-Cosine and scores are normalized using the s-norm method. The full description of this method can be found in~\cite{zeinali2017hmm}.
	
	
	
	\section{Evaluation Protocol}
	
	\subsection{Challenge Website}
	
	The challenge has a Github page (\url{https://sdsvc.github.io/}). All information and news about the challenge will be posted on the website. System description submitted by the participants will be made available on the website as well.
	The Github page of SdSV Challenge 2020 was completely moved to \url{https://sdsvc.github.io/2020/}.
	
	\subsection{Leaderboard platform}
	
	As described above, there is an online leaderboard for each task and participants can submit maximum 10 systems per day during the period of evaluation. The leaderboard shows the performance of the systems on the final evaluation set (unlike SdSVC 2020).
	The challenge leaderboard platforms are available at
	\begin{itemize}
		\item Task 1:~ \url{https://competitions.codalab.org/competitions/28190}
		\item Task 2:~ \url{https://competitions.codalab.org/competitions/28269}
	\end{itemize}

	\subsection{System Output Format}
	
	The results should be submitted is a single ZIP file that contains two files: answer.txt and metadata. The files should be at the root of the ZIP file and the ZIP file should not contain any folders and other unwanted files.

    The answer.txt is a one-column text file. Each line of it indicates a LLR score (a float number) of the corresponding trial. The order of scores must be the same as the file of the trials and all of the trials must be scored. Any inconsistency will cause an error in the evaluation of the system. Please note that there is a header line in the trial file but the header should not be included in the answer.txt file.
    
	For example:
	\begin{verbatim}
	-6.1284
	-97.8528
	-16.8025
	-44.3276
	4.4121
	-61.0123
	-42.9890
	\end{verbatim}

    The metadata is a text file that contains information about the system you are submitting in the following format:

    \begin{itemize}
        \item public-description: a short description of the submitted method. This will be added to the detailed results page.
        \item fused-systems-count: an integer number that indicates the number of fused systems for achieving this submission.
    \end{itemize}

    An example of metadata for the baseline system can be:
    
    {\em
    public-description: This is a submission by SdSV Challenge organizers as a baseline. It is based on the standard x-vector recipe for speaker verification.
    
    fused-systems-count: 1
	}

    If the files are not found inside the ZIP file or if it is not correctly formatted as specified above, your submission will most likely fail and it will NOT be scored. However, note that improper formatting might also lead to the wrong interpretation of your results, and therefore to the WRONG score.
	
	\subsection{Data License Agreement}
	
	The evaluation data for this challenge is a subpart of the DeepMine dataset. To use this dataset, participants should sign a data license agreement specific to SdSV 2021 Challenge if they did not participate in SdSV Challenge 2020. This license allows participants to use the data for the challenge as well as the subsequent paper publication. Any other usage of the data is not allowed. The license agreement file can be found on the challenge website.
	
	\subsection{System Description}
	
	Each participant is required to submit a full system description. The system description will be made available online. 
	In addition to the system description, we strongly recommend participants to submit papers to the special session related to the challenge in InterSpeech2021. The papers will be reviewed as a normal paper, so they should be in a proper format and has sufficient novelty for acceptance.
	
	The system description should have at least 2 pages and must include the following information about the submitted systems:
	
	\begin{itemize}
		\item a complete description of the system components, including front-end and back-end modules along with their configurations.
		\item a complete description of the data partitions used to train the various models.
		\item performance of the submitted systems on the progress and evaluation subsets reported on the leaderboard website.
	\end{itemize}
	
	Bibtex source for citing the evaluation plan is provided below for easy reference.
	
	\begin{verbatim}
	@techreport{sdsvc2021plan,
	    title={Short-duration Speaker Verification (SdSV) Challenge 2021: 
	    the Challenge Evaluation Plan.},
	    author={Zeinali, Hossein and Lee, Kong Aik and Alam, Jahangir and 
	    Burget, Luka\v{s}},
	    institution={arXiv preprint arXiv:1912.06311},
	    year=2020,
	}
	\end{verbatim}
	
	\section{Planned Evaluation Schedule}
	
	\begin{table}[h]
		\renewcommand{\arraystretch}{1.2}
		\centerline
		{
			\setlength\tabcolsep{20pt}
			\begin{tabular}{l | l }
				Release of evaluation plan:				& Jan 10, 2021 \\
				Evaluation platform open:				& Jan 15, 2021 \\
				Release of Train, Dev, and Eval sets:	& Jan 15, 2021 \\
				Challenge deadline:						& Mar 20, 2021 \\
				INTERSPEECH Paper submission:			& Mar 26, 2021 \\
				System description deadline:			& Apr 06, 2021 \\
				SdSV Challenge 2021 special session at INTERSPEECH:				& Aug 20 - Sep 03, 2021 \\
			\end{tabular}
		}
	\end{table}
	
	\bibliographystyle{IEEEbib}
	\bibliography{refs}
	
\end{document}